\documentstyle[nato,psfig,numreferences]{crckapb}

\begin{opening}
\title{GENERALIZED DISTRIBUTION AMPLITUDES: NEW TOOLS TO STUDY
HADRONS' STRUCTURE AND INTERACTION}
\subtitle{Talk presented at
the Second International "Cetraro" Workshop $\&$ NATO Advanced
Research Workshop "Diffraction 2002", Alushta, Crimea, Ukraine,
August 31 - September 6, 2002. }
\author{R. Fiore}
\institute{Dipartimento di Fisica, Universit\`a della Calabria $\&$ \\
INFN-Cosenza, I-87036 Arcavacata di Rende, Cosenza, Italy
}
\author{A. Flachi}
\institute{IFAE, Universidad Aut\`onoma de Barcelona\\
08193 Bellaterra, Barcelona, Spain
}
\author{L.L.~Jenkovszky}
\institute{N.N. Bogolyubov Institute for Theoretical Physics\\
Academy of Sciences of Ukraine\\
Metrogohichna 14b, 03143 Kiev, Ukraine}
\author{A. Lengyel}
\institute{Institute of Electron Physics\\
Universitetska 21, UA-88000 Uzhgorod, Ukraine
}
\author{\underline{V.K.~Magas}}
\institute{N.N. Bogolyubov Institute for Theoretical Physics\\
Academy of Sciences of Ukraine\\
Metrogohichna 14b, 03143 Kiev, Ukraine $\&$ \\
Center for Physics of Fundamental
Interactions (CFIF)\\
Physics Department, Instituto Superior Tecnico\\
Av. Rovisco Pais, 1049-001 Lisbon, Portugal}

\end{opening}

\runningtitle{GENERALIZED DISTRIBUTION AMPLITUDES: NEW TOOLS TO STUDY
HADRONS' STRUCTURE AND INTERACTION}

\newcommand{\beq}{\begin{equation}}
\newcommand{\eeq}[1]{\label{#1} \end{equation}}
\newcommand{\insertplot}[1]{\centerline{\psfig{figure={#1},width=13.0cm}}}

\newcommand{\insertplotlll}[1]{\centerline{\psfig{figure={#1},height=18.0cm}}}

\begin{document}

% The \begin{document} command comes after the \end{opening}
% command.

\begin{abstract}
A non-perturbative approach to Generalized Parton Distributions, and to
Deeply Virtual Compton Scattering in particular, based on off mass shell
extension of dual amplitudes with Mandelstam analyticity (DAMA) is
developed with the spin and helicity structure as well as the threshold
behavior accounted for. The model is tested against the data on deep
inelastic electron-proton scattering from JLab.
\end{abstract}

\section{Introduction} \label{s1}

Parton distributions measure the probability that a quark or gluon
carry a fraction $x$ of the hadron momentum. Relevant structure functions (SF)
are related by unitarity to the imaginary part of the forward Compton
scattering amplitude. Generalized parton distributions (GPD) \cite{M,Ji,Rad}
represent the
interference of different wave functions: one, where a parton carries momentum
fraction $x+\xi$, and the other one, with momentum
fraction $x-\xi$, correspondingly; here $\xi$ is skewedness and can
be determined by the external momenta in a deeply virtual Compton scattering
(DVCS) experiment. Basically, the DVCS amplitude can be viewed as a  binary
hadronic scattering amplitude continued off mass shell.  DVCS  is
related to the GPD in the same way as elastic Compton scattering
is related to the ordinary SFs.

Apart from the longitudinal momentum fraction of gluons
and quarks, GPDs contain also information about their transverse location,
given by a Fourier transformation over the $t$-dependent GPD. Real-space
images
of the target can thus be obtained in a completely new way \cite{Pire}.
Spacial resolution is determined by the virtuality of the incoming photon.
Quantum photographs of the nucleons and  nuclei with resolutions on the scale of a
fraction of femtometer are thus feasible.

One of the first experimental observations of DVCS was based on the recent
analysis of the JLab data from the CLAS collaboration. New
measurements at higher energies are currently being analyzed, and dedicated
experiments are planned \cite{Elou}.

\begin{figure}[htb]
%\vspace*{7.5cm}
        \insertplot{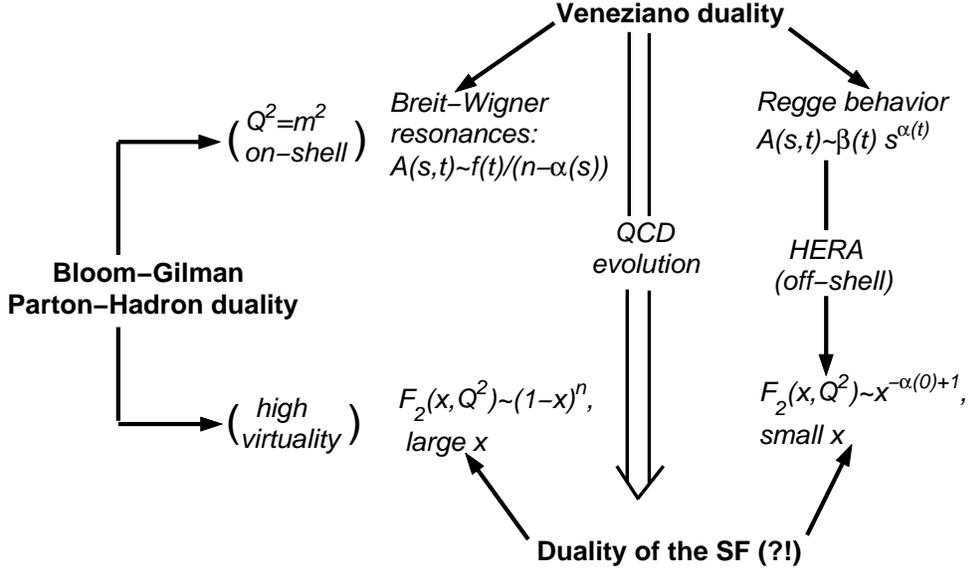}
%\vspace*{-0.5cm}
\caption[]{Generalized parton distribution amplitudes. A
road map.}
%\vspace*{-0.6cm}
\label{diag}
\end{figure}

On the theoretical side, much progress has been achieved
\cite{M,Ji,Rad} in treating GPD in the framework of quantum chromodynamics (QCD)
and the light-cone technique. On the other hand,
non-nonpertubative effects
(resonance production, the background, low-$Q^2$ effects)
dominating the kinematical region of present measurements and the underlying
dynamics still leave much ambiguity in the above-mentioned
field-theoretical approach.
Therefore, as an alternative or complementary approach we have suggested
\cite{JMP,JMP-1,FFJLM} to use dual amplitudes with
Mandelstam analyticity (DAMA) as a model for GPD in general and DVCS in
particular. We remind that DAMA realizes duality between direct-channel
resonances and high-energy Regge behavior (``Veneziano-duality''). By
introducing $Q^2$-dependence in DAMA, we have extended the model off mass
shell and have shown
\cite{JMP,JMP-1} how parton-hadron (or ``Bloom-Gilman'') duality
is realized in this
way. With the above specification, DAMA can serve as and explicit model
valid for
all values of the Mandelstam variables $s$, $t$ and $u$ as well as any $Q^2$,
thus
realizing the ideas of DVCS and related GPDs. The historical and logical
connections between different kinametical regions and relevant models are
depicted in Fig. \ref{diag}.

In this work
we concentrate on the very delicate and disputable problem of the off mass shell
continuation (introduction of the variable $Q^2$) in the dual model,
starting with inclusive electron-nucleon scattering
%shown in Fig.\ref{rat},
both at high energies, typical of HERA, and low energies, with the
JLab data in mind (see ref. \cite{FFJLM} for more details).

\section{Simplified model}
\label{s22}

The central object of the present study is the nucleon SF,
uniquely related to the photoproduction cross section by
\beq
F_2(x,Q^2)={Q^2(1-x)\over{4\pi \alpha (1+{4m^2 x^2\over
{Q^2}})}} \sigma_t^{\gamma^*p}(s,Q^2)\ ,
\eeq{m23}
where the total
cross section, $\sigma_t^{\gamma^* p}$, is the imaginary part of
the forward Compton scattering amplitude, $A(s, Q^2)$,
$\sigma_t^{\gamma^* p}(s)={\cal I}m\ A(s, Q^2)$;
$m$ is the nucleon mass, $\alpha$ is the fine structure constant.
 The center of mass energy of the $\gamma^* p$ system,
 the negative squared photon virtuality $Q^2$ and the Bjorken
 variable $x$ are related by
$s=Q^2(1-x)/x+m^2$.

We adopt the two-component picture of strong interactions
\cite{FH}, according to which direct-channel resonances are dual
to cross-channel Regge exchanges and the smooth background in the
$s-$channel is dual to the Pomeron exchange in the $t-$channel.
This nice idea, formulated \cite{FH} more than three decades ago, was first
realized explicitly in the framework of DAMA with nonlinear trajectories
(see \cite{JKM,JMP} and earlier references therein).

 As explained in Refs.~\cite{JMP} and \cite{JKM}, the background in a dual
model corresponds
to a pole term with an exotic trajectory that does not
produce any resonance.

In the dual-Regge approach \cite{JMP,JMP-1,FFJLM}  Compton
scattering can be viewed as an off mass shell continuation of a
hadronic reaction, dominated in the resonance region by
direct-channel non-strange ($N$ and $\Delta$) baryon trajectories.
The scattering
amplitude follows from the pole decomposition of a dual amplitude
\cite{JMP}
\beq
A(s,Q^2)\Biggl|_{t=0}=
norm\sum_{i=N_1^*,N_2^*,\Delta,E}A_i
\sum_{n=n_i^{min}}^{n_i^{max}}{f_{i}(Q^2)^{2\left(n-n_i^{min}+1\right)}
\over{n-\alpha_{i}(s)}}\ ,
\eeq{m17}
where $i$ runs over all the
trajectories allowed by quantum number exchange, $norm$ and
$A_i$'s are constants, $f_i(Q^2)$'s are the form factors. These
form factors generalize the concept of inelastic (transition) form
factors to the case of continuous spin, represented by the
direct-channel trajectories. The $n_i^{min}$ refers to the spin of
the first resonance on the corresponding trajectory $i$ (it is
convenient to shift the trajectories by $1/2$, therefore we use
$\alpha_i=\alpha^{phys}_i-1/2$, which due to the semi-integer
values of the baryon spin leaves $n$ in Eq.~(\ref{m17}) integer).
The sum over $n$ runs with step $2$ (in order to conserve parity).

It follows from Eq.~(\ref{m17}) that
\beq
{\cal I}m\
A(s,Q^2)=norm\sum_{i=N_1^*,N_2^*,\Delta,E}A_i
\sum_{n=n_i^{min}}^{n_i^{max}}{[f_i(Q^2)]^{2\left(n-n_i^{min}+1\right)}
{\cal I}m\ \alpha_i(s) \over {(n-{\cal R}e\
\alpha_i(s))^2+\left({\cal I}m\ \alpha_i(s)\right)^2}}\ .
\eeq{sumsum}

\begin{table}[ht]
\caption[]{Values of the fitted parameters. In the first column we show the
result of the fit when the parameters of the baryonic trajectories
are fixed. The second column contains the result of the fit when
the parameters of the trajectories are  varied. $^\dagger$ denotes
the parameters of the physical baryon trajectories from ref.
\cite{FFJLM}. \linebreak
 $^*$ The coefficient $norm$ is chosen in such
a way as to keep $A_{N_1^*}=1$ in order to see the interplay
between different resonances.\linebreak
$^\diamond$ Using intercepts and thresholds as a free parameters does not improve the fit, but
they may get values far from original.} \vspace{0.5cm}
\begin{tabular}{|c|c|c|c|c|}
\hline
  &$\alpha_{0}$                   & -0.8377 (fixed)$^\dagger$  & -0.8070 &
  -0.8377 (fixed)$^\diamond$\\
  &$\alpha_{1}$                   &  0.95  (fixed)$^\dagger$  &  0.9632 & 0.9842\\
$N_1^*$           & $\alpha_{2}$  &   0.1473 (fixed)$^\dagger$ & 0.1387 &  0.1387\\
  &$A_{N_1^*}$                    &  1 (fixed)$^*$&  1 (fixed)$^*$ & --\\
  &$Q^2_{N_1^*}$, GeV$^2 $                  &  2.4617& 2.6066 & --\\
\hline
  &$\alpha_{0}$                   &  -0.37(fixed)$^\dagger$  & -0.3640 &
  -0.37(fixed)$^\diamond $\\
  &$\alpha_{1}$                   &   0.95  (fixed)$^\dagger$ &  0.9531 & 0.9374 \\
$N_2^*$           & $\alpha_{2}$  &   0.1471 (fixed)$^\dagger$ & 0.1239 & 0.1811\\
 & $A_{N_2^*}$                   &  0.5399& 0.6086 & -- \\
  & $Q^2_{N_2^*}$, GeV$^2 $                 &   2.9727& 2.6614  & -- \\
\hline
  &$\alpha_{0}$                   &  0.0038 (fixed)$^\dagger$  & -0.0065  &
  0.0038 (fixed)$^\diamond$\\
  &$\alpha_{1}$                   &  0.85    (fixed)$^\dagger$ & 0.8355 & 0.8578\\
$\Delta$          & $\alpha_{2}$  &  0.1969  (fixed)$^\dagger$ &  0.2320 & 0.2079\\
  & $A_{\Delta}$                  &   4.2225&  4.7279 & -- \\
  & $Q^2_{\Delta}$, GeV$^2 $                &   1.5722 & 1.4828  & -- \\
\hline
  & $s_{0}$, GeV$^2 $                    &  1.14 (fixed)$^\dagger$   & 1.2871 &
  1.14 (fixed)$^\diamond$\\
\hline
  &$\alpha_{0}$                   &  0.5645  &  0.5484 & 7.576 \\
  & $\alpha_{2}$  &  0.1126    & 0.1373  & 0.0276 \\
${E}$   & $s_{E}$, GeV$^2 $                    &  1.3086  & 1.3139  & 1.311 (fixed)$^\diamond$\\
 & $A_{exot}$                    &  19.2694  &  14.7267 & -- \\
  & $G_{exot}$                    &  --  &  -- & 24.16 \\
  & $Q^2_{exot}$, GeV$^2 $                  &  4.5259  &  4.6041 &  4.910 \\
\hline
DS & $Q_0^2$ & -- & -- & 2.691 \\
 & ${Q^{\prime}_0}^2$ & -- & -- & 0.4114 \\
\hline
& $norm$ & 0.021 & 0.0207 & 0.0659 \\
\hline
&  $\chi^2_{d.o.f.}$                 &  28.29    & 11.60  & 18.1 \\
\hline
\end{tabular} \label{t1}
\end{table}

The first three terms in (\ref{sumsum}) are the non-singlet, or
Reggeon contributions with the $N^*$ and $\Delta$  trajectories in
the  $s$-channel, dual to the exchange  of an effective bosonic
trajectory (essentially, $f$) in the $t$-channel, and the fourth
term is the contribution from the smooth background, modeled by a
non-resonance pole term with an exotic trajectory $\alpha_E(s)$,
dual to the Pomeron (see Ref.~\cite{JMP}). As argued in
Ref.~\cite{JMP}, only a limited number, ${\cal N}$, of resonances
appear on the trajectories, for which reason we tentatively set
${\cal N}=3$ - one resonance on each trajectories ($N^*_1,\
N^*_2,\ \Delta$), i.e. $n_i^{max}=n_i^{min}$.
Our analyses \cite{FFJLM} shows that ${\cal
N}=3$ is a reasonable approximation.
The limited (small) number of resonances
contributing to the cross section results not only from the
termination of resonances on a trajectory but even more due to the
strong suppression coming from the numerator (increasing powers of
the form factors).

We use Regge trajectories with a threshold
singularity and nonvanishing imaginary part in the form: \beq
\alpha(s)=\alpha_0+\alpha_1 s+\alpha_2(\sqrt {s_0}-\sqrt{s_0-s}),
\eeq{m18} where $s_0$ is the lightest threshold,
$s_0=(m_{\pi}+m_p)^2=1.14$ GeV$^2$ in our case, and the linear term
approximates the contribution from heavy
thresholds \cite{JMP,JMP-1,FFJLM}.

For the exotic trajectory we also keep only one term in the sum \footnote{In Ref.
\cite{JKM} the whole DAMA integral was calculated numerical
 and it has been shown that in the resonance region
the direct-channel exotic trajectory gives a non-negligible contribution,
amounting to about 10-12$\%$.}. $n_E^{min}$ is the first integer
larger then $Max({\cal R}e\ \alpha_E)$ -- to make sure that there are no
resonances on the exotic trajectory.
The exotic trajectory is taken in the form
\begin{equation}
\alpha_E(s)=\alpha_E(0)+\alpha_{1E}(\sqrt{s_E}-\sqrt{s_E-s})\ ,
\end{equation}
where $\alpha_E(0)$, $\alpha_{1E}$ and the effective
exotic threshold $s_E$ are free parameters.

To start with, we  use the simplest, dipole model for the form
factors, disregarding the spin structure of the amplitude and the
difference between electric and magnetic form factors:
\beq
f_i(Q^2)=\left(1+{Q^2/Q^2_{0,i}}\right)^{-2}\ ,
\eeq{ff} where
$Q_{0,i}^2$ are scaling parameters, determining the relative
growth of the
resonance peaks and background.

%\pagebreak
\begin{figure}[ht]
%\vspace*{-1.0cm}
        \insertplotlll{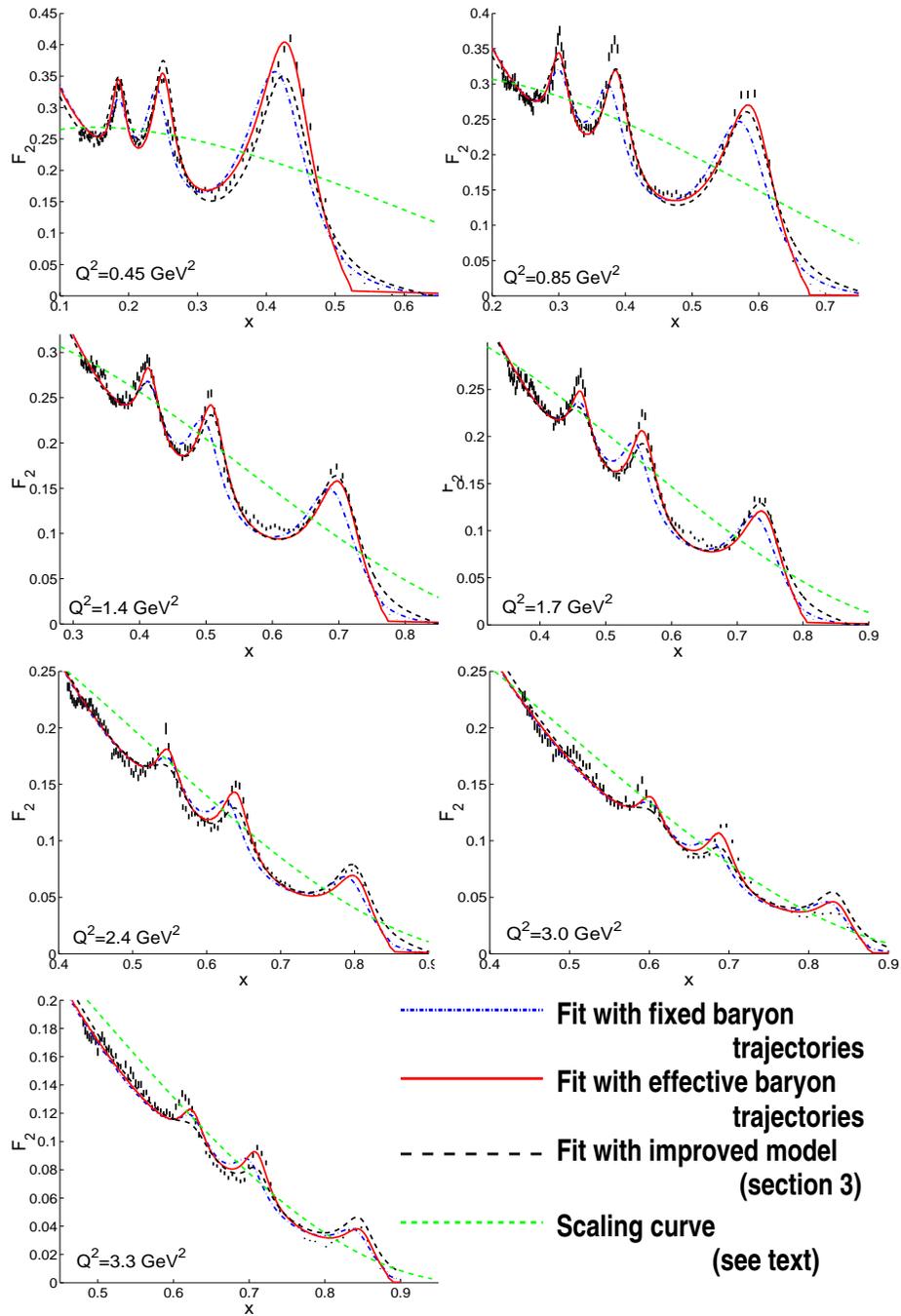}
\vspace*{-0.5cm} \caption{$F_2$ as a function of $x$ for $Q^2 =
0.45-3.3$ GeV$^2$. } \label{fit1}
\end{figure}

We test our model against the experimental data
from SLAC \cite{Stoler} and JLab
\cite{SLAC}\footnote{We are grateful to M.I. Niculescu for making
her data compilation available to us.}.
This set of the experimental data is not
homogeneous, i.e. points at low $s$ (high $x$) are given with
very small experimental errors, thus ``weighting'' the fitting
procedure not uniformly. Consequently, a
preselection of the date involved in the fitting procedure was made,
although we present all
the experimental
points in the Figure. The results of
our fits and the values of the fitted parameters are presented in Figs. \ref{fit1}
and Table \ref{t1}. More details
can be found in ref. \cite{FFJLM}.

For the first fit - first column of Table \ref{t1} and dashed-dotted lines in Fig. \ref{fit1} -
we fix the parameters of the $\Delta$ and $N^*$ trajectories to the physical ones,
reproducing the correct masses and widths of the resonances, leaving the four scaling
constants $Q^2_i$, four factors $A_i$ and the parameters of the exotic
trajectories to be fitted to the data.
Then, to improve the model we effectively account for the large number
of overlapping resonances (about 20) present in the energy range under investigation.
For this reason we consider the dominant resonances
($N^*_1$, $N^*_2$ and $\Delta$) as ``effective'' contributions to the SF.
In other words, we require that they mimic the contribution of
the dominant resonances plus the large number of subleading
contributions, which, together, fully describe the real physical system.
In the light of these considerations, we have refitted the data,
allowing the baryon trajectories parameters to vary. The resulting
parameters of such a fit
are reported  in Table \ref{t1} (second column).
It is worth noting that although the range of variation was not restricted,
the new parameters of the trajectories stay close to their physical values,
showing stability of the fit and thus reinforcing our previous considerations.
From the relevant plots, shown in Fig. \ref{fit1} with full lines,
one can see that
the improvement is significant, although agreement is still far from being
perfect ($\chi^2_{d.o.f.}=11.6$).

The smooth dashed lines in Fig. (\ref{fit1}) correspond to a
``scaling curves'', i.e. a phenomenological
parameterizations of the SF exhibiting Bjorken scaling and
fitting the data \cite{BCDMS}.

\section {Spin and helicity; threshold behaviour}
\label{s6}

In the preceding section the nucleon form factor was treated in a
phenomenological way, fitted to the data, neglecting the longitudinal cross
section $\sigma_L$, and the $Q^2\rightarrow 0$ limit of the SFs.
On the other hand, it is
known \cite{DS,CM} that with account for both $\sigma_L$ and
$\sigma_T$ the form factor can be presented as sum o three terms:
$G_+(Q^2)$, $G_0(Q^2)$ and $G_0(Q^2),$ corresponding to $\gamma^*
N\rightarrow R$ helicity transition amplitudes in the rest frame
of the resonance $R$:
\beq
G_{\lambda_\gamma}={1\over
m}<R,\lambda_R=\lambda_N-\lambda_{\gamma}|J(0)|N,\lambda_N>;
\eeq{m16}
here $\lambda_R, \lambda_N$ and
$\lambda_\gamma$ are the resonance, nucleon and photon helicities,
$J(0)$ is the current operator; $\lambda_\gamma$ assumes the
values $-1, 0$ and $+1.$ Correspondingly, one replaces the
$Q^2-$dependent expression in the numerator of (\ref{sumsum}) by
\beq
f(Q^2)^N \rightarrow|G_+(Q^2)|^2+2|G_0(Q^2)|^2+|G_-(Q^2)|^2.
\eeq{m15}

The explicit form of these form factors is known only near their
thresholds $|\vec q |\rightarrow 0$, while their large-$Q^2$
behavior may be constrained by the quark counting rules.

    According to \cite{BW}, one has near the threshold
\beq
G_{\pm}(Q^2) \sim|\vec q|^{J-3/2}, \ \ G_0(Q^2)
\sim{q_0\over{|\vec q|}}|\vec q|^{J-1/2}
\eeq{thr1}
for the so-called normal ($1/2^+\rightarrow 3/2^-, 5/2^+,
7/2^-,...$) transitions and
\beq
G_{\pm}(Q^2 \sim|\vec q|^{J-1/2}, \ \ G_0(Q^2)
\sim{q_0\over{|\vec q|}}|\vec q|^{J+1/2}
\eeq{thr2}
for the anomalous ($1/2^+\rightarrow 1/2^-, 3/2^+, 5/2^-,...$)
transitions, where
\beq
|\vec q|={\sqrt{(M^2-m^2-Q^2)^2+4M^2Q^2}\over{2M}}, \ \ |\vec
q|_{Q=0}={M^2-m^2\over {2M}},
\eeq{m12}
\beq
q_0 = \frac{M^2-m^2-Q^2}{2M}\ ,
\eeq{q0}
$M$ is a resonance mass \footnote{In our approach $M$ is defined by trajectory:
${\cal R}e \alpha(M^2)=J$, where $J$ is a spin of the resonance;
and therefore for the varying baryon trajectory
$M$ might differ from the mass of physical resonance.}.

 Following the quark counting rules, in refs. \cite{CM} (for a
 recent treatment see
\cite {DS}), the large-$Q^2$ behavior of $G$'s was assumed to be
\beq
G_+(Q^2)\sim Q^{-3}, \ \ G_0(Q^2)\sim Q^{-4},\ \ G_-(Q^2)\sim
Q^{-5}.
\eeq{QCR}
Let us note that while this is reasonable (modulo logarithmic
factors) for elastic form factors, it may not be true any more for
inelastic (transition) form factors. For example,  dual models
(see Eq. (1) and ref. \cite {JMP}) predict powers of the form
factors to increase with increasing excitation (resonance spin).
This discrepancy can be resolved only experimentally, although a
model-independent analysis of the $Q^2$-dependence for various
nuclear excitations is biased by the (unknown) background.

In ref. \cite{DS} the following expressions for the $G$'s,
combining the above threshold- (\ref{thr1}), (\ref{thr2}) with the
asymptotic behavior (\ref{QCR}), was suggested:
\beq
|G_{\pm}|^2=|G_{\pm}(0)|^2\Biggl({|\vec q|\over{|\vec q|_{Q=0}}}
{Q'^2_0\over{Q^2+Q_0'^2}}\Biggr)^{2J-3}
\Biggl({Q'^2_0\over{Q^2+Q_0'^2}}\Biggr)^{m_{\pm}}
\eeq{m10}

\beq
|G_0|^2=C^2
\Biggl({Q^2_0\over{Q^2+Q_0^2}}\Biggr)^{2a}
{q_0^2\over{|\vec q|^2}}
\Biggl({|\vec q|\over{|\vec q|_{Q=0}}}{Q'^2_0\over{Q^2+Q_0'^2}}\Biggr)^{2J-1}
\Biggl({Q^2_0\over{Q^2+Q_0^2}}\Biggr)^{m_0}
\eeq{m9}
for the normal transitions and

\beq
|G_{\pm}|^2=|G_{\pm}(0)|^2\Biggl({|\vec q|\over{|\vec q|_{Q=0}}}
{Q'^2_0\over{Q^2+Q_0'^2}}\Biggr)^{2J-1}
\Biggl({Q'^2_0\over{Q^2+Q_0'^2}}\Biggr)^{m_{\pm}}
\eeq{m8}

\beq
|G_0|^2=C^2
\Biggl({Q^2_0\over{Q^2+Q_0^2}}\Biggr)^{2a}
\Biggl({q_0^2\over{|\vec q|^2}} {Q'^2_0\over{Q^2+Q_0'^2}}\Biggr)^{2J+1}
\Biggl({Q^2_0\over{Q^2+Q_0^2}}\Biggr)^{m_0}
\eeq{m7}
for the anomalous ones, where $m_+=3$, $m_0=4$, $m_-=5$ count the quarks,
$C$ and $a$ are free parameters. The form factors at $Q^2=0$
are related to the known (measurable) helicity photoproduction
amplitudes $A_{1/2}$ and $A_{3/2}$ by
\beq
|G_{+,-}(0)|=\frac{1}{\sqrt{4\pi\alpha}}\sqrt{M\over{M-m}}|A_{1/2,3/2}|.
\eeq{m6}
The values of the helicity amplitudes are quoted by
experimentalists \cite{PDG} (those relevant to the present
discussion are compiled also in \cite{DS}).

We have fitted the above model to the JLab data \cite{SLAC} again by
keeping the contribution from three prominent resonances, namely
$\Delta(1232)$, $N^*(1520)$ and $N^*(1680)$. For the sake of simplicity
we neglected the cross term containing $G_0$ (and the coefficient
$C$ and parameter $a$), since it is small relative to the other two
terms in Eq. (\ref{m15}) \cite{DS}.

We use the same background term as in the previous section, but with the
normalization coefficient $G_E$:

$${background}={{f_E}{I_E}\over{(n_E^{min}-R_{E})^2}+I_{E}^2}\ ,\ \
{f_E}= G_E\Biggl({{Q^2_E}\over{Q^2+Q^2_E}}\Biggr)^4\ .$$

To be specific, we write explicitly the three resonance terms, to
be fitted to the data, (cf. with Eq. (\ref{sumsum}) of the previous
section)\footnote{Remember that our trajectories are shifted
by $1/2$ from the physical trajectories
$\alpha_i=\alpha^{phys}_i-1/2$.}:
$${\cal I}m\ A(s,Q^2)=norm\cdot$$
\beq
{f_{\Delta}I_{\Delta}\over{(1-R_{\Delta})^2+I_{\Delta}^2}}+
{f_{N^-}I_{N^-}\over{(1-R_{N^-})^2+I_{N^-}^2}}+
{f_{N^+}I_{N^+}\over{(2-R_{N^+})^2+I_{N^+}^2}}+ background,
\eeq{m4}
where e.g. $f_{\Delta}$ is calculated according to Eqs. (\ref{m15},\ref{m10},\ref{m8}):
\beq
f_{\Delta}=\Biggl({|\vec q|\over{|\vec q|_{Q=0}}}
{Q'^2_0\over{Q^2+Q_0'^2}}\Biggr)^{2J-1=2}
\Biggl(|G_+(0)|^2\Bigl({Q^2_0\over{Q^2+Q_0^2}}\Bigr)^3+
|G_-(0)|^2\Biggl({Q^2_0\over{Q^2+Q_0^2}}\Bigr)^5\Biggr);
\eeq{m3}
here $R$ and $I$ denote the real and the imaginary parts of the
relevant trajectory, specified by the subscript.
Similar expressions can be easily cast for $f_{N^+}$ and for
$f_{N^-}$ as well.

The form factors at $Q^2=0$ can be simply calculated from Eq. (\ref{m6})
by inserting the known \cite{PDG} (see also Table 1 in
\cite {DS}) values of the relevant photoproduction amplitudes:

$A_{\Delta(1232)}(1/2, 3/2) = (-0.141\ GeV^{-1/2}, -0.258\
GeV^{-1/2});$

$ A_{N(1520))} (1/2, 3/2) = (-0.022\ GeV^{-1/2}, 0.167\ GeV^{-1/2})$;

$A_{N(1680)} (1/2, 3/2) = (-0.017\ GeV^{-1/2}, 0.127\ GeV^{-1/2}).$

Note that in this way the relative normalization of the resonance terms is
fixed (no $A_i$'s appearing in previous section), leaving only two adjustable parameters,
$Q_0^2$ and $Q'^2_0$ (provided $C$ is set zero, then $a$ also disappears),
which means that this version
of the model (with the cross term, containing $G_0$, neglected!)
is very restrictive.

The resulting fits to the SLAC and JLab data are presented in
Fig. \ref{fit1} (dashed lines) and Table \ref{t1} (third column).
The fit is not so good ($\chi^2_{d.o.f.}=18.1$), what probably tells
us not to neglect the cross term.
We hope to account for this and to improve the fit in the forthcoming works.

\section {Future Prospects}\label{s9}

In this work we have presented new results on the extension of a
dual model (Sec. 2) to include the spin and helicity structure of the
amplitudes (SFs) as well as its threshold behavior as $Q^2 \rightarrow 0$.
Let us remind that the lowest threshold (in the direct channel) of a hadronic
reaction (if Compton scattering is considered in analogy with $\pi N$
scattering) is $s_0=(m_\pi+m_N)^2$. Hence its imaginary part and the relevant
structure function starts at this value and vanishes below $s_0$. The
``gap'' between the lowest electromagnetic $m_p^2$ and the above-mentioned
hadronic thresholds is filled by the arguments and formulas of Sec. \ref{s6}.
The $Q^2\rightarrow 0$ is also important as the
transition point between elastic and inelastic dynamics.

The important step performed in our work after Ref. \cite{DS} is the
``Reggization'' of the Breit-Wigner pole terms (\ref{m4}), i.e.  single
resonance terms in Ref. \cite{DS} are replaced by those including relevant
baryon trajectories\footnote{Another important step is adding the background
into consideration.}.
The form of these trajectories, constrained by analyticity, unitarity and
by the experimental data is crucial for
the dynamic. The use of baryon trajectories instead of
individual resonances not only makes the model economic (several
resonances are replaced by one trajectory) but also helps in
classifying the resonances, by including the ``right'' ones and
eliminating those nonexistent. The construction of feasible models of baryon
trajectories, fitting
the data on resonances and yet satisfying the theoretical bounds, is in
progress.

Note also that the dual model constrains the form factors:
as seen from Eq. (\ref{sumsum}), the powers of the form factors rise with increasing
spin of the excited state. This property of the model can and should be
tested experimentally.

The $Q^2$ dependence, introduced in this model via the form factors can be
studied and constrained also by means of the QCD evolution equations.

To summarize, the model presented in this paper, can be used as a laboratory
for testing various ideas of the analytical S-matrix and quantum chromodynamics.
Its virtue is a simple explicit form, to be elaborated and constrained
further both by theory and experimental tests.

{\it Acknowledgment.} L.J., A.L. and V.M. acknowledge the support
by INTAS, grants 00-00366 and 97-31726.

\vfill \eject

\end{document}